\documentclass[letterpaper,journal]{IEEEtran}

\usepackage{mathtools,amssymb,amsfonts,bm}
\usepackage{array}
\usepackage{textcomp}
\usepackage{mathabx}
\usepackage{booktabs}
\usepackage{multirow}

\usepackage[caption=false,font=normalsize,labelfont=sf,textfont=sf]{subfig}
\usepackage{stfloats}
\usepackage{graphicx}
\usepackage{adjustbox}
\usepackage[flushleft]{threeparttable}

\usepackage{url}
\usepackage[noadjust]{cite}
\usepackage[ruled,lined,longend,linesnumbered]{algorithm2e}
\usepackage{xcolor}
\usepackage{microtype}

\definecolor{skyblue}{RGB}{135,206,235}

\newcommand{\figplaceholder}[2][0.24\textheight]{%
  \fbox{\parbox[c][#1][c]{0.95\linewidth}{\centering\textbf{Figure placeholder}\\[2pt]\texttt{\detokenize{#2}}}}%
}
\newcommand{\safeincludegraphics}[2][]{%
  \IfFileExists{#2}{\includegraphics[#1]{#2}}{%
    \IfFileExists{fig/#2}{\includegraphics[#1]{fig/#2}}{\figplaceholder{#2}}%
  }%
}

\newcommand{\defeq}{\coloneqq}

\newcommand{\setB}{\mathcal B}
\newcommand{\setD}{\mathcal D}
\newcommand{\setG}{\mathcal G}
\newcommand{\setR}{\mathcal R}

\newcommand{\setPQ}{\setD}

\newcommand{\setref}{\setR}
\newcommand{\setPQPV}{\setB\!\setminus\!\setR}
\newcommand{\setPVref}{\setB\!\setminus\!\setD}

\newcommand{\nB}{n_{\mathcal B}}
\newcommand{\nD}{n_{\mathcal D}}
\newcommand{\nG}{n_{\mathcal G}}
\newcommand{\nR}{n_{\mathcal R}}

\newcommand{\nBatch}{N_{\mathrm{batch}}}
\newcommand{\nLim}{N_{\mathrm{lim}}}
\begin{document}

\title{SABLE: GPU-Based Power Flow Accelerator for \\ Sparsity-Aware Batched Learning}

\author{Suho Park, Keunju Song, and Hongseok Kim, \IEEEmembership{Senior Member, IEEE}}

\maketitle

\begin{abstract}
Recent studies have developed GPU-based approaches for solving AC power flow and successfully applied them to standalone power flow problems. However, integrating these approaches into modern differentiable learning frameworks while preserving sparsity remains challenging. To this end, we present SABLE, a GPU-based sparse batched power flow accelerator for differentiable learning via an implicit power flow layer.
SABLE leverages a block-diagonal embedding that
reformulates batched three-dimensional Jacobians as a fixed-pattern two-dimensional sparse template that is shared across PyTorch,
CuPy, and cuDSS. This common template enables zero-copy
interoperability and memory-efficient sparse reuse across the software stack. On top of this representation, SABLE accelerates repeated power flow computations through reusable sparse templates, custom GPU kernels, a cuDSS-based sparse-direct LU solver, and mixed-precision techniques. Extensive experiments show that SABLE improves standalone power flow solving throughput by up to $253.4\times$ over \emph{pandapower} and $5.7\times$ over \emph{\emph{ExaPF}}. In end-to-end training, evaluated on AC optimal power flow learning models based on DC3 and DeepLDE, SABLE expands the feasible training batch range by up to $64\times$ and improves training throughput by up to $206.7\times$ over the corresponding baseline.
\end{abstract}


\section{Introduction}
\IEEEPARstart{T}{he} increasing penetration of distributed energy resources (DERs) is shifting grid security assessment from deterministic snapshot analysis to high-dimensional, scenario-driven analysis. Applications such as probabilistic power flow \cite{prusty2017critical}, large-scale uncertainty sweeps, and data generation for machine learning require solving thousands to tens of thousands of AC power flow (PF) instances at low latency. This demand has motivated substantial research on accelerating batched PF evaluation. 

However, these advances have been aimed primarily at standalone PF workloads, while research on batched differentiable implicit PF layers remains limited. Consequently, existing physics-embedded learning systems such as DC3 and DeepLDE still lack a sparsity-preserving PF execution path capable of sustaining high throughput under repeated training-time use \cite{donti2021dc3, kim2025deepl}.
They therefore rely primarily on dense, memory-intensive GPU computation which makes the PF layer a dominant bottleneck in both runtime and memory during end-to-end training. The challenge is thus no longer merely to solve PF quickly once, but to execute it repeatedly, sparsely, and differentiably within large-scale learning loops.

These requirements expose three coupled limitations of prior work: a structural limitation in differentiable sparse integration, a performance limitation in repeated Jacobian and linear equation solving executions, and FP64-centric PF solving on modern GPU architecture.

\subsection{Structural Barriers to Reusable Sparse Differentiable PF}

Existing studies on PF acceleration have reported
substantial speedups. However, these methods were developed
primarily to maximize standalone PF solving throughput, often
through custom kernels implemented in performance-oriented
environments such as C++ or Julia rather than within the
Python-based deep-learning stack \cite{sooknanan2016gpu, exapf_juliacon, garcia2012parallel, li2014gpu_epsr, zhou2017gpu, zhou2017tps}. Even when Python bindings
are available, they typically provide only a wrapper-level
interface, while the core sparse computation is not naturally
embedded in the learning framework and its computational graph \cite{wang2021segan}.
This creates a structural gap between efficient standalone PF
execution and the end-to-end differentiation required when solving PF
is invoked repeatedly during training.

This gap becomes more pronounced in large-scale batched
training. Traditional batched PF implementations often assume
a 3D batched sparse representation to process many scenarios
simultaneously \cite{zhou2017gpu ,zhou2017tps, wang2021segan}. Such design is natural in standalone batched PF solving, but it does not transfer cleanly to PyTorch sparse
execution. PyTorch sparse support does not readily match a
direct 3D batched sparse workflow; sparse compressed tensors
are still defined with only two sparse dimensions, slicing
support is limited, and the available sparse operations and
backward paths cover only a narrow set of cases \cite{pytorch_sparse_doc}. As a result, directly mapping 3D batched PF layers into PyTorch does
not readily provide a reusable sparse execution path for
repeated forward and backward passes.

A new architecture is therefore needed to embed batched PF
layers into the training loop while preserving
differentiability and enabling efficient reuse of sparse
structure. Such architecture should also minimize
format-conversion overhead and manage GPU memory efficiently
so that large-scale training can be sustained without
sacrificing the computational advantages of sparse PF
execution.

\subsection{Performance Limits of Repeated Jacobian and Sparse-Direct Execution}

Beyond the structural limitations discussed above, there is
also a performance limitation. Early PF acceleration studies were concerned primarily with
reducing the cost of solving a single PF problem. Because the graph structure of power systems is highly sparse, most prior work exploited
sparsity-aware techniques and was developed mainly on CPU
platforms, where sparse and irregular computations were more
manageable \cite{cui2021effective, ahmadi2022fast}. These studies were largely based on the Newton-Raphson (NR) method and therefore focused on its two main
computational bottlenecks: Jacobian construction and the solution
of the associated linearized equations. In doing this, previous work
investigated multicore-parallel Jacobian computation \cite{ahmadi2022fast, kuyumcu2024efficient}, LU-based
sparse-direct solvers such as KLU \cite{su2020full}, and algorithms tailored to
multicore execution \cite{cui2021effective}. 

As the system size and the number of target scenarios increased, however, CPU-based approaches became limited in
throughput, which necessitated GPU-based PF acceleration. Hence, Jacobian computation on GPU evolved from single-scenario
acceleration to batched acceleration of many scenarios \cite{sooknanan2016gpu, exapf_juliacon, wang2021segan}. Similarly, linear-system solution techniques also progressed from iterative methods \cite{garcia2012parallel, li2014gpu_epsr} to more effective sparse LU-based approaches for batched PF \cite{zhou2017gpu, zhou2017tps, wang2021segan}.

Nevertheless, these GPU-oriented studies have not yet been fully adapted to physics-informed neural networks \cite{donti2021dc3, kim2025deepl}. Specifically, the accelerator is required to be embedded in training, and sparse representation should remain compatible with the learning process while preserving memory efficiency during repeated training and NR iterations. Furthermore, the linear equation solving stage should be compatible with updating the Jacobian for each training sample. This requires both faster PF solving and a
reusable sparse numerical path for repeated Jacobian refresh so that
linear equation solution resides in the learning loop.

\subsection{Performance Limits of Double-Precision-Centric PF on Modern GPUs}

As discussed above, GPUs are essential for batched PF
acceleration, both for standalone PF and implicit PF layer. Their main benefit is high-throughput parallel
processing across many scenarios. However, this benefit is not
easy to realize for PF and other numerically rigorous problems; such problems have traditionally relied on double-precision (DP) arithmetic
because of their stringent numerical requirements \cite{thurner2018pandapower}.

This creates a performance limitation on modern GPUs. On
architectures such as A100 and H100, standard single-precision (SP)
throughput is already about $2\times$ higher than DP
throughput. The gap is much larger on Tensor Core paths.
Reduced-precision modes provide roughly $8$--$16\times$
higher throughput on A100 and about $14$--$30\times$ higher
throughput on H100, relative to DP Tensor Core execution \cite{nvidia_a100_spec, nvidia_h100_spec}.
As a result, a DP-dominant PF execution path is not well
matched to the throughput profile of modern GPUs. This mismatch becomes more severe in PF layer embedded in learning loop because PF should be executed repeatedly across
NR iterations as well as forward and backward
passes during training. A GPU-efficient differentiable PF
layer therefore needs a mixed-precision (MP) strategy that preserves numerical robustness while reducing DP exposure in the dominant numerical computations.

To overcome the three aforementioned limitations, we present a novel framework called SABLE, which stands for Sparsity-aware Accelerator for a Batched Learning Engine. As the name suggests, SABLE is GPU-based and devised for differentiable learning via an implicit PF layer. The main contributions are summarized as follows.

\begin{itemize}
    \item \textbf{Shared sparse template for differentiable batched PF:}
    SABLE introduces a block-diagonal embedding (BDE) that reformulates batched PF Jacobians into a single fixed-pattern two-dimensional sparse template. BDE is reused throughout training, and this shared template enables the GPU-resident differentiable implicit PF layer for both standalone PF evaluation and forward/backward execution. Built on top of the BDE structure, SABLE also supports zero-copy interoperability across the libraries used in training, thereby avoiding redundant memory allocation.
    
    \item \textbf{Fixed-pattern Jacobian reuse and sparse-direct execution:}
    On the shared BDE template, SABLE initializes the batched Jacobian sparsity pattern and symbolic analysis state only once. SABLE further develops an in-place value-update kernel that refreshes only the nonzero Jacobian entries. A block-aware indexing kernel then extracts the Jacobian operators needed for forward and backward passes, while cuDSS-based sparse-direct LU refactorization and solve reuse the unchanged symbolic structure across repeated iterations. This avoids redundancies such as repeated sparse-pattern construction, symbolic reanalysis, and per-iteration memory reallocation.
    
    \item \textbf{Stagewise MP for solving PF using the Newton-Raphson method:}
    Instead of using DP throughout the entire NR procedure, SABLE retains DP for the outer nonlinear logic while executing the linear equation solving in SP. This improves performance while preserving numerical robustness.
\end{itemize}

The paper is organized as follows. Section II formulates the embedded PF layer and its underlying standalone PF formulation. Section III presents SABLE, focusing on a sparsity-preserving and learning-compatible execution path and custom kernel/mixed-precision strategies for accelerated PF computation. Section IV evaluates SABLE in standalone PF and embedded PF layer training settings. Section V concludes.

\section{Problem Formulation}
\label{sec:problem}
We formulate a unified batched PF problem setting that covers both standalone batched PF evaluation and its use as a differentiable implicit PF layer in training.

\subsection{Bus-Type Notation and AC Power Flow Equations}
\label{subsec:pf_nr_formulation}

Let $\setD$, $\setG$, and $\setR$ denote the load (PQ), generator (PV), and
slack bus index sets, with cardinalities $\nD$, $\nG$, and $\nR$, respectively.
Let $\setB=\setD\cup\setG\cup\setR$ denote the set of all buses, and let
$\nB=\nD+\nG+\nR$. We consider a balanced AC transmission network with admittance
matrix $\bm Y_{\mathrm{bus}}=\bm G+j\bm B\in\mathbb C^{\nB\times\nB}$. Let \(\nBatch\) denote the batch size, and each index
\(b=1,\ldots,\nBatch\) corresponds to one PF scenario. The set \(\{\bm u^{(b)}\}_{b=1}^{\nBatch}\) represents scenario-dependent operating-point data, while all scenarios share the same topology and bus-type partition.
For each scenario $b$, the complex voltage at bus $i$ is
$V_i^{(b)}=V_{m,i}^{(b)}e^{j\theta_i^{(b)}}$. The specified net injections are
$P_{i,\mathrm{spec}}^{(b)}=p_{g,i}^{(b)}-p_{d,i}^{(b)}$ and
$Q_{i,\mathrm{spec}}^{(b)}=q_{g,i}^{(b)}-q_{d,i}^{(b)}$. At a PF solution, the
calculated injections $P_i^{(b)}$ and $Q_i^{(b)}$ match these specified values. However, mismatches occur until convergence and are denoted by $\Delta P_i^{(b)}=P_{i,\mathrm{spec}}^{(b)}-P_i^{(b)}$ and $\Delta Q_i^{(b)}=Q_{i,\mathrm{spec}}^{(b)}-Q_i^{(b)}$.~\cite{Tinney1967}.

\subsection{Embedded PF Layer Formulation with Standalone Evaluation}
\label{subsec:completion_formulation}

Following \cite{donti2021dc3,kim2025deepl}, we formulate the embedded PF
layer using equality completion in the \textit{forward} pass and implicit differentiation in the \textit{backward} pass. Note that the same forward completion also serves as a standalone PF evaluator.
Given the scenario data $\bm u^{(b)}$, we define the full PF buffer used to assemble the completed PF solution and the full PF mismatch vector, as
\begin{equation}
\begin{aligned}
\bm y^{(b)}
&\defeq
\begin{bmatrix}
\bm p_{g,\setPVref}^{(b)}\\
\bm q_{g,\setPVref}^{(b)}\\
\bm V_m^{(b)}\\
\bm \theta^{(b)}
\end{bmatrix},
&
\bm F^{(b)}(\bm y^{(b)};\bm u^{(b)})
&\defeq
\begin{bmatrix}
\Delta \bm P^{(b)}\\
\Delta \bm Q^{(b)}
\end{bmatrix}.
\end{aligned}
\label{eq:y_and_F_base}
\end{equation}
Our vector notation uses set-specific subscripts. For example, $\bm p_{g,\setPVref}^{(b)}$ denotes the active power of generators at all buses excluding load buses. We omit the bus-set subscript when it is for
all-bus quantities such as $\bm V_m^{(b)}$ and $\bm\theta^{(b)}$.
In $\bm y^{(b)}$, generation coordinates at PQ buses are omitted because
their active and reactive generations are fixed at zero by the bus-type
definition. The slack-bus angle $\bm\theta_{\setref}^{(b)}$ is retained
in $\bm\theta^{(b)}$ only for notational consistency and remains fixed
as the reference angle. Under these bus-type conventions, the forward
pass enforces the AC PF equalities by satisfying
$\bm F^{(b)}(\bm y^{(b)};\bm u^{(b)})=\bm 0$.

The corresponding \textit{base} Jacobian is
\begin{equation}
\bm J^{(b)}
\defeq
\frac{\partial \bm F^{(b)}}{\partial \bm y^{(b)}}
=
\begin{bmatrix}
\dfrac{\partial \Delta\bm P}{\partial \bm p_{g,\setPVref}} &
\dfrac{\partial \Delta\bm P}{\partial \bm q_{g,\setPVref}} &
\dfrac{\partial \Delta\bm P}{\partial \bm V_m} &
\dfrac{\partial \Delta\bm P}{\partial \bm\theta} \\[8pt]
\dfrac{\partial \Delta\bm Q}{\partial \bm p_{g,\setPVref}} &
\dfrac{\partial \Delta\bm Q}{\partial \bm q_{g,\setPVref}} &
\dfrac{\partial \Delta\bm Q}{\partial \bm V_m} &
\dfrac{\partial \Delta\bm Q}{\partial \bm\theta}
\end{bmatrix}^{(b)} .
\label{eq:J_base_block}
\end{equation}
This Jacobian defines the common sparse template for three different Jacobians for the forward and backward passes. We will see that this enables the cached indexing strategy in
Section~III, where each operator is refreshed from the shared template rather
than separately reconstructed.

\subsubsection{Forward pass with Equality Completion}
\label{subsubsec:forward pass completion}

Based on this PF representation, the neural network supplies only the
partial decision block
\begin{equation}
\bm x^{(b)}
:=
\begin{bmatrix}
\bm p^{(b)}_{g,\mathcal G}\\
\bm V^{(b)}_{m,\mathcal B\setminus\mathcal D}
\end{bmatrix}.
\label{eq:ynn}
\end{equation}
Note that when $\bm x^{(b)}$ is populated with case-specified values and
$\bm u^{(b)}$ provides the load data of a standalone PF case, this process can also be used for a standalone batched PF evaluation. The completed
PF buffer $\bm y^{(b)}$ is then formed by assembling three parts, i.e., the neural network output $\bm x^{(b)}$, the
NR-completed block $\bm z_1^{(b)}$, and the explicit-recovery block
$\bm z_2^{(b)}$, up to the fixed slack-angle coordinate:
\begin{align}
\bm z_1^{(b)}
&:=
\begin{bmatrix}
\bm V^{(b)}_{m,\mathcal D}\\
\bm \theta^{(b)}_{\mathcal B\setminus\mathcal R}
\end{bmatrix},
\label{eq:z1_block}
\\
\bm z_2^{(b)}
&:=
\begin{bmatrix}
\bm p^{(b)}_{g,\mathcal R}\\
\bm q^{(b)}_{g,\mathcal B\setminus\mathcal D}
\end{bmatrix}.
\label{eq:z2_block}
\end{align}
\noindent\textbf{Step~1.} Given the neural output $\bm x^{(b)}$ and the scenario data $\bm u^{(b)}$,
the forward pass first solves the \textit{traditional} PF having $\bm V^{(b)}_{m,\mathcal D}$, $\bm \theta^{(b)}_{\mathcal B\setminus\mathcal R}$ in \eqref{eq:z1_block} for NR completion equations
\begin{equation}
\bm F_1^{(b)}
\left(
\bm z_1^{(b)};
\bm x^{(b)},\bm u^{(b)}
\right)
\defeq
\begin{bmatrix}
\Delta \bm P_{\setPQPV}^{(b)}\\[2pt]
\Delta \bm Q_{\setPQ}^{(b)}
\end{bmatrix}
=
\bm 0 .
\label{eq:reduced_pf_system}
\end{equation}
At the $t$-th NR iteration, this gives
\begin{equation}
\bm J_1^{(b)}\!\left(\bm z_{1,t}^{(b)}\right)
\delta\bm z_{1,t}^{(b)}
=
-\bm F_1^{(b)}
\left(
\bm z_{1,t}^{(b)};
\bm x^{(b)},\bm u^{(b)}
\right),
\label{eq:nr_linear_system_completion}
\end{equation}
followed by the update
$\bm z_{1,t+1}^{(b)}=\bm z_{1,t}^{(b)}+\delta\bm z_{1,t}^{(b)}$. The Step~1 Jacobian is evaluated at the current NR iteration:
\begin{equation}
\bm J_1^{(b)}\!\left(\bm z_{1,t}^{(b)}\right)
\defeq
\begin{bmatrix}
\dfrac{\partial \Delta\bm P_{\setPQPV}}{\partial \bm V_{m,\setPQ}} &
\dfrac{\partial \Delta\bm P_{\setPQPV}}{\partial \bm\theta_{\setPQPV}} \\[10pt]
\dfrac{\partial \Delta\bm Q_{\setPQ}}{\partial \bm V_{m,\setPQ}} &
\dfrac{\partial \Delta\bm Q_{\setPQ}}{\partial \bm\theta_{\setPQPV}}
\end{bmatrix}^{(b)}_{\bm z_{1,t}} .
\label{eq:step1_jacobian}
\end{equation}
\noindent\textbf{Step~2.} Once $\bm z_1^{(b)}$ is obtained, the remaining variables of $\bm y^{(b)}$, i.e., $\bm z_2^{(b)}$, are recovered from
the unused equality rows:
\begin{equation}
\bm F_2^{(b)}
\left(
\bm z_2^{(b)};
\bm z_1^{(b)},\bm x^{(b)},\bm u^{(b)}
\right)
\defeq
\begin{bmatrix}
\Delta \bm P_{\setref}^{(b)}\\[2pt]
\Delta \bm Q_{\setPVref}^{(b)}
\end{bmatrix}
=
\bm 0 .
\label{eq:F2_completion}
\end{equation}
Since $\bm z_2^{(b)}$ enters $\bm F_2^{(b)}$ additively with unit coefficients,
$\partial \bm F_2^{(b)}/\partial \bm z_2^{(b)}=\bm I$, and Step~2 reduces to a straightforward explicit recovery of the remaining variables.

\subsubsection{Backward pass with Implicit Differentiation}

Now, all derivatives are evaluated at the converged completion solution. We
focus here on the formulation of the two key Jacobian operators that determine
the computational flow of SABLE and on their associated numerical operations. 
Let $\mathcal L$ denote the scalar loss function. Since the forward completion first solves the NR block $\bm z_1^{(b)}$ and then recovers the remaining equality-completion block $\bm z_2^{(b)}$, the total derivative with respect to the neural output $\bm x^{(b)}$ follows the two-step completion chain rule:
\begin{equation}
\begin{aligned}
\frac{d\mathcal L}{d\bm x^{(b)}}
&=
\frac{\partial\mathcal L}{\partial \bm x^{(b)}}
+
\frac{\partial\mathcal L}{\partial \bm z_1^{(b)}}
\frac{\partial \bm z_1^{(b)}}{\partial \bm x^{(b)}} \\
&\quad+
\frac{\partial\mathcal L}{\partial \bm z_2^{(b)}}
\frac{\partial \bm z_2^{(b)}}{\partial \bm x^{(b)}}
+
\frac{\partial\mathcal L}{\partial \bm z_2^{(b)}}
\frac{\partial \bm z_2^{(b)}}{\partial \bm z_1^{(b)}}
\frac{\partial \bm z_1^{(b)}}{\partial \bm x^{(b)}} .
\end{aligned}
\label{eq:chain_rule}
\end{equation}

First, let $\bm v^{(b)} := [\bm V_m^{(b)};\bm\theta^{(b)}]$ denote the
voltage-coordinate block. The Step~2 voltage-coordinate Jacobian is then
\begin{equation}
\bm J_2^{(b)}
:=
\frac{\partial \bm F_2^{(b)}}{\partial \bm v^{(b)}}
=
\begin{bmatrix}
\dfrac{\partial \Delta \bm P_R^{(b)}}{\partial \bm V_m^{(b)}} &
\dfrac{\partial \Delta \bm P_R^{(b)}}{\partial \bm\theta^{(b)}} \\
\dfrac{\partial \Delta \bm Q_{\mathcal B\setminus\mathcal D}^{(b)}}{\partial \bm V_m^{(b)}} &
\dfrac{\partial \Delta \bm Q_{\mathcal B\setminus\mathcal D}^{(b)}}{\partial \bm\theta^{(b)}}
\end{bmatrix}.
\label{eq:Jstep2_impl}
\end{equation}

Since $\partial \bm F_2^{(b)}/\partial \bm z_2^{(b)}=\bm I$, it follows that $\partial \bm z_2^{(b)}/\partial \bm v^{(b)}=-\bm J_2^{(b)}$.
Therefore, the Step~2 gradient contribution on the voltage-coordinate
block is computed by one transpose sparse matrix--vector multiplication (SpMV):
\begin{equation}
\bar{\bm v}_2^{(b)}
:=
-
\left(\bm J_2^{(b)}\right)^\top
\left(
\frac{\partial\mathcal L}{\partial \bm z_2^{(b)}}
\right)^\top ,
\end{equation}

which provides, after
transposition, the two chain rule terms in \eqref{eq:chain_rule},
\[
\frac{\partial\mathcal L}{\partial \bm z_2^{(b)}}
\frac{\partial \bm z_2^{(b)}}{\partial \bm z_1^{(b)}}
\quad \text{and} \quad
\frac{\partial\mathcal L}{\partial \bm z_2^{(b)}}
\frac{\partial \bm z_2^{(b)}}{\partial \bm x^{(b)}} .
\]

Second, we collect the direct and Step~2-induced contributions into the
intermediate row-gradients
\begin{equation}
\bm g_{z_1}^{(b)}
:=
\frac{\partial\mathcal L}{\partial \bm z_1^{(b)}}
+
\frac{\partial\mathcal L}{\partial \bm z_2^{(b)}}
\frac{\partial \bm z_2^{(b)}}{\partial \bm z_1^{(b)}},
\qquad
\bm g_x^{(b)}
:=
\frac{\partial\mathcal L}{\partial \bm x^{(b)}}
+
\frac{\partial\mathcal L}{\partial \bm z_2^{(b)}}
\frac{\partial \bm z_2^{(b)}}{\partial \bm x^{(b)}} .
\end{equation}
Here, $\bm g_{z_1}^{(b)}$ and $\bm g_x^{(b)}$ are the accumulated
row-gradients after the Step~2 recovery has been propagated backward. To propagate $\bm g_{z_1}^{(b)}$ through Step~1 without forming $\partial \bm z_1^{(b)}/\partial \bm x^{(b)}$, the Step~1 Jacobian used in the forward NR completion is \textit{reused} in transposed form:
\begin{equation}
\left(\bm J_1^{(b)}\right)^\top \bm\lambda^{(b)}
=
\left(\bm g_{z_1}^{(b)}\right)^\top ,
\end{equation}
where $\bm\lambda^{(b)}$ is the Step~1 auxiliary vector. The Step~1
Jacobian block with respect to the voltage-magnitude input is
\begin{equation}
\bm J_3^{(b)}
:=
\frac{\partial \bm F_1^{(b)}}{\partial \bm V_{m,\mathcal B\setminus\mathcal D}^{(b)}}
=
\begin{bmatrix}
\dfrac{\partial \Delta \bm P_{\mathcal B\setminus\mathcal R}^{(b)}}%
{\partial \bm V_{m,\mathcal B\setminus\mathcal D}^{(b)}} \\[1.2ex]
\dfrac{\partial \Delta \bm Q_{\mathcal D}^{(b)}}%
{\partial \bm V_{m,\mathcal B\setminus\mathcal D}^{(b)}}
\end{bmatrix}.
\label{eq:Jstep3_impl}
\end{equation}

Combining the selector operation for the active-power input with the
transpose SpMV associated with $\bm J_3^{(b)}$, the final
gradient with respect to the neural input block is
\begin{equation}
\left(
\frac{d\mathcal L}{d\bm x^{(b)}}
\right)^\top
=
\left(\bm g_x^{(b)}\right)^\top
-
\begin{bmatrix}
\bm\lambda_G^{(b)} \\
\left(\bm J_3^{(b)}\right)^\top \bm\lambda^{(b)}
\end{bmatrix},
\end{equation}
where $\bm\lambda_G^{(b)}$ denotes the entries of $\bm\lambda^{(b)}$
associated with the $\Delta P$ rows on PV buses.

\noindent\textbf{Connection to the proposed method.}
This formulation reveals that training the whole passes repeatedly, in both forward and backward, requires the forward Jacobian $\bm J_1$ while $\bm J_1^\top$, $\bm J_2^\top$, and $\bm J_3^\top$ are
required for the backward pass. Without a reusable sparse
structure, these operators are repeatedly reconstructed, incurring memory reallocation overhead at every NR iteration and training step, which is the current bottleneck in the implicit PF layer. In addition to this overhead, the PF layer is dominated by sparse linear system solving and transpose SpMVs. Therefore, it requires a GPU-resident framework that reuses the fixed sparse pattern, updates only numerical values, and connects these operations efficiently to the training loop. Section III presents SABLE to realize this execution path.

\section{Proposed Method}
\label{sec:proposed}

We now describe the proposed SABLE engine that realizes the formulation in Section~\ref{sec:problem} on GPUs. SABLE consists of three components: \emph{(i)} a learning-compatible sparse execution path for the PF layer built from BDE and zero-copy interoperability, \emph{(ii)} one-time symbolic analysis followed by repeated numeric execution on fixed sparse structures, and \emph{(iii)} stagewise mixed precision.

\subsection{Learning-Compatible Sparse Execution via BDE and Zero-Copy Interoperability}
\label{subsec:interop_BDE}

The formulation in Section~II indicates that PF layer execution entails repeated operations involving sparse Jacobian operators. In this setting, a 3D batched sparse representation becomes a structural
bottleneck for reusable sparse execution in modern learning frameworks. SABLE overcomes this bottleneck through BDE while preserving sparsity and numerical equivalence. Building on this BDE representation, SABLE further
establishes a zero-copy interoperability framework across PyTorch, CuPy, and cuDSS to maximize GPU-resident memory efficiency.

\begin{figure}[!t]
    \centering
    \safeincludegraphics[width=1\linewidth]{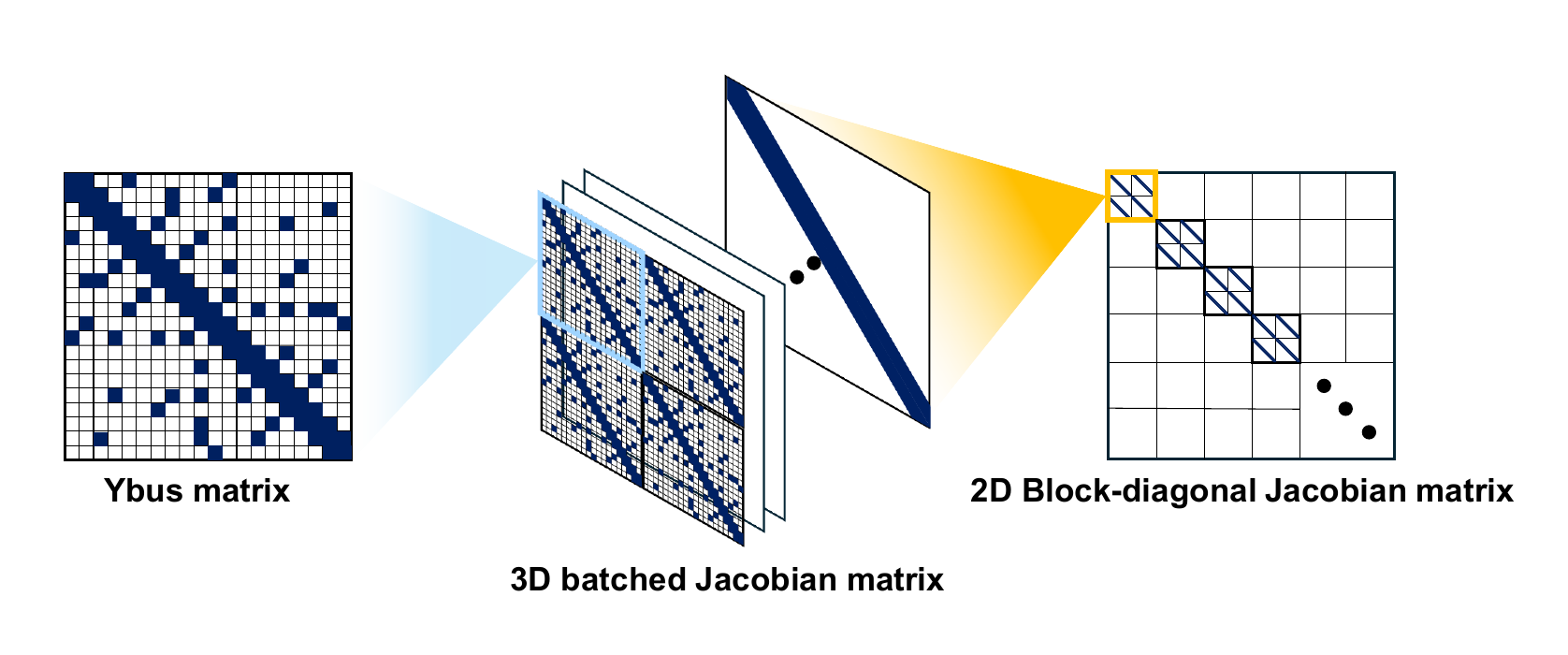}
    \caption{BDE: from a single-scenario sparsity pattern to a batch-global fixed 2D template.}
    \label{fig:fig_SABLE_PA_BDE}
\end{figure}

\subsubsection{Block-Diagonal Embedding (BDE)}
Let \(\bm A^{(b)}\in\mathbb{R}^{M\times N}\) denote an operator for scenario \(b\),
covering the base Jacobian \(\bm J^{(b)}\) in~\eqref{eq:J_base_block} and the
Jacobian operators extracted from it. For each operator type, all scenarios
share the same network topology; hence, the corresponding operators share
the same sparsity pattern, defined by the row--column index pairs
\(\Omega=\{(r_k,c_k)\}_{k=1}^{K}\), where \(K\) is the number of nonzeros (nnz) in each \(\bm A^{(b)}\).

Fig.~\ref{fig:fig_SABLE_PA_BDE} illustrates the overall BDE procedure for
constructing a 2D sparse template from a 3D batched sparse representation. Specifically, BDE embeds these batched sparse operators into
a single 2D global sparse matrix \(\bm A_{\mathrm{gl}}\) by placing them on
the block diagonal:
\begin{equation}
\bm{A}_{\mathrm{gl}}
=\bigoplus_{b=1}^{\nBatch}\bm{A}^{(b)}
=\operatorname{diag}(\bm{A}^{(1)},\dots,\bm{A}^{(\nBatch)}).
\label{eq:BDE_diag}
\end{equation}
Accordingly, the global sparsity pattern is obtained by applying row and column offsets to the single-scenario pattern:
\begin{equation}
\begin{aligned}
\Omega_{\mathrm{gl}}
&=
\bigcup_{b=1}^{\nBatch}
\bigl\{(r_k+(b-1)M,\;c_k+(b-1)N) \\
&\qquad\mid (r_k,c_k)\in\Omega\bigr\}.
\end{aligned}
\label{eq:BDE_global}
\end{equation}
These resulting global sparse pattern is constructed only once for a given
topology and batch size, and are reused across NR iterations and training
steps.

This embedding enables the batched operators to be handled as one sparse
matrix without altering the independent computation of each scenario.
Applying this equivalence to the \(t\)-th NR iteration in
\eqref{eq:nr_linear_system_completion} with
\(\bm A^{(b)}=\bm J_{1,t}^{(b)}\) gives
\begin{equation}
\begin{aligned}
\bm J_{1,t,\mathrm{gl}}\delta\bm z_{1,t,\mathrm{gl}}
&=
-\bm F_{1,t,\mathrm{gl}} \\
&\Longleftrightarrow\quad
\bm J_{1,t}^{(b)}\delta\bm z_{1,t}^{(b)}
=
-\bm F_{1,t}^{(b)},\;\; b=1,\dots,\nBatch .
\end{aligned}
\label{eq:BDE_forward_equiv}
\end{equation}
where the global correction and residual vectors are formed by stacking each per-scenario vector as
\[
\delta\bm z_{1,t,\mathrm{gl}}
=
\begin{bmatrix}
\delta\bm z_{1,t}^{(1)}\\
\vdots\\
\delta\bm z_{1,t}^{(\nBatch)}
\end{bmatrix},
\quad
\bm F_{1,t,\mathrm{gl}}
=
\begin{bmatrix}
\bm F_{1,t}^{(1)}\\
\vdots\\
\bm F_{1,t}^{(\nBatch)}
\end{bmatrix}.
\]

Consequently, unlike previous learning-compatible PF layers that rely on dense tensor representations, BDE provides the structural basis for
sparsity-preserving end-to-end training while maintaining numerical
equivalence.

\begin{figure*}[t]
    \centering
    \safeincludegraphics[width=\textwidth]{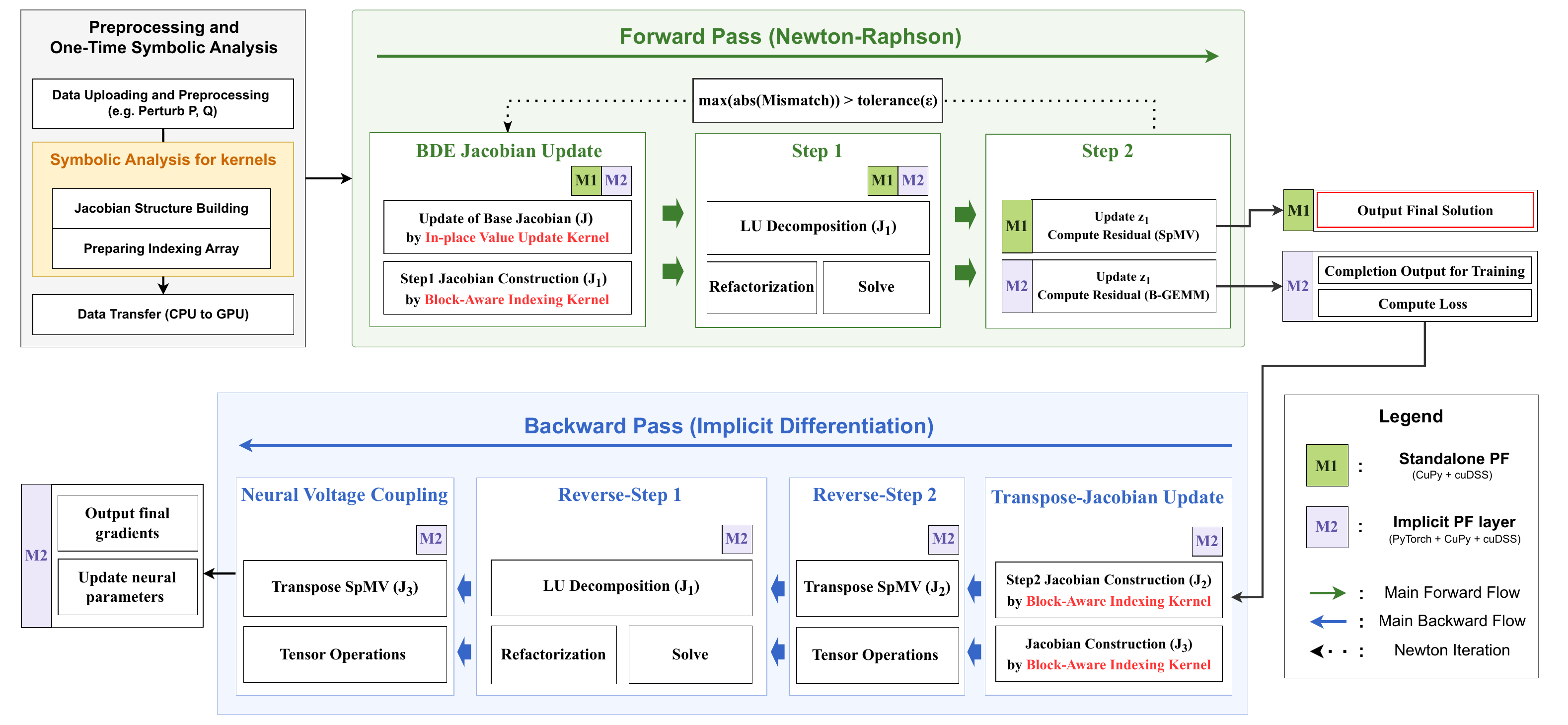}
    \caption{Overall execution flow of SABLE for standalone PF and training with an implicit PF layer.}
    \label{fig:wide}
\end{figure*}

\subsubsection{DLPack Interchange and Device-Pointer Binding}
Fig.~\ref{fig:wide} summarizes the overall structure of SABLE. M1 denotes the standalone PF path built on CuPy and cuDSS via nvmath-python, whereas M2 denotes the training path built on PyTorch, CuPy, and the same cuDSS interface via nvmath-python. Across both paths, end-to-end memory efficiency depends on keeping the BDE sparsity structure fixed from start to finish and updating only the numerical values. This avoids repeated sparse-memory reallocation and structural reconstruction, which would otherwise introduce unnecessary overhead.

To realize this design, SABLE uses two forms of zero-copy GPU interoperability. First, PyTorch and CuPy exchange dense state vectors, indexing arrays, and BDE Jacobian matrices in coordinate (COO) format through DLPack, so the same device-resident buffers can be shared without copying~\cite{pytorch_dlpack_doc}. The COO matrices are represented by \texttt{row}, \texttt{column}, and \texttt{value} buffers. Second, the linear solving stage is connected through nvmath-python, specifically the low-level \texttt{nvmath.bindings.cudss} interface. For this solver-facing stage, the required BDE Jacobians are maintained in compressed sparse row (CSR) format, represented by \texttt{indptr}, \texttt{indices}, and \texttt{values} buffers, which are bound to cuDSS matrix objects through their CUDA device pointers together with the dense right-hand-side and solution buffers~\cite{nvidia_nvmath_python}.

As a result, the sparse template and associated workspaces are instantiated once and reused across iterations, while only Jacobian values and related dense vectors are refreshed in place. This keeps the full BDE-based pipeline GPU-resident and memory-efficient in both standalone (M1) and learning-loop (M2) settings.

\begin{algorithm}[t]
\caption{In-Place Value-Update Kernel}
\label{alg:kernel_fill}
\KwIn{
Batch size $\nBatch$;
number of kernel-updated entries $K_{\mathrm{upd}}$;
single-scenario Jacobian nnz $K_J$;
cached metadata
$\{p_k,\mathrm{blk}_k,i_k,j_k,G_k,B_k\}_{k=1}^{K_{\mathrm{upd}}}$;
refreshed state arrays
$V_m,V_{\mathrm{re}},V_{\mathrm{im}},\theta,I_{\mathrm{re}},I_{\mathrm{im}}$;
preallocated global value buffer
$\bm v\in\mathbb{R}^{\nBatch K_J}$
}
\KwOut{The same buffer $\bm v$ with updated entries}

\For{$b=1,\ldots,\nBatch$ and $k=1,\ldots,K_{\mathrm{upd}}$ \tcp*[f]{in parallel}}{
    Read $(p,\mathrm{blk},i,j,G,B)\leftarrow
    (p_k,\mathrm{blk}_k,i_k,j_k,G_k,B_k)$\;

    Read local state values
    $(v_{m,i},v_{m,j},v_{\mathrm{re},i},v_{\mathrm{im},i},
    \iota_{\mathrm{re},i},\iota_{\mathrm{im},i},\theta_i,\theta_j)
    \leftarrow
    (V_{m,i}^{(b)},V_{m,j}^{(b)},V_{\mathrm{re},i}^{(b)},
    V_{\mathrm{im},i}^{(b)},I_{\mathrm{re},i}^{(b)},
    I_{\mathrm{im},i}^{(b)},\theta_i^{(b)},\theta_j^{(b)})$\;

    \uIf{$\mathrm{blk}=dP_{\theta}$}{
        $\eta\leftarrow
        \begin{cases}
        v_{\mathrm{im},i}\iota_{\mathrm{re},i}
        -v_{\mathrm{re},i}\iota_{\mathrm{im},i}
        +Bv_{m,i}^{2},
        & i=j\\
        -v_{m,i}v_{m,j}
        \left(G\sin\theta_{ij}-B\cos\theta_{ij}\right),
        & i\ne j
        \end{cases}$\;
    }
    \uElseIf{$\mathrm{blk}=dP_{V_m}$}{
        $\eta\leftarrow
        \begin{cases}
        -\dfrac{
        v_{\mathrm{re},i}\iota_{\mathrm{re},i}
        +
        v_{\mathrm{im},i}\iota_{\mathrm{im},i}
        }{v_{m,i}}
        -Gv_{m,i},
        & i=j\\
        -v_{m,i}
        \left(G\cos\theta_{ij}+B\sin\theta_{ij}\right),
        & i\ne j
        \end{cases}$\;
    }
    \uElseIf{$\mathrm{blk}=dQ_{\theta}$}{
        $\eta\leftarrow
        \begin{cases}
        -\left(
        v_{\mathrm{re},i}\iota_{\mathrm{re},i}
        +
        v_{\mathrm{im},i}\iota_{\mathrm{im},i}
        \right)
        +Gv_{m,i}^{2},
        & i=j\\
        v_{m,i}v_{m,j}
        \left(G\cos\theta_{ij}+B\sin\theta_{ij}\right),
        & i\ne j
        \end{cases}$\;
    }
    \uElseIf{$\mathrm{blk}=dQ_{V_m}$}{
        $\eta\leftarrow
        \begin{cases}
        -\dfrac{
        v_{\mathrm{im},i}\iota_{\mathrm{re},i}
        -
        v_{\mathrm{re},i}\iota_{\mathrm{im},i}
        }{v_{m,i}}
        +Bv_{m,i},
        & i=j\\
        -v_{m,i}
        \left(G\sin\theta_{ij}-B\cos\theta_{ij}\right),
        & i\ne j
        \end{cases}$\;
    }

    $\pi\leftarrow (b-1)K_J+p$\;
    $\bm v[\pi]\leftarrow\eta$\;
}
\end{algorithm}

\subsection{One-Time Symbolic Analysis and Repeated Numeric Execution}
\label{subsec:symbolic_numeric}

Once the learning-compatible and memory-efficient structure is established, SABLE performs symbolic setup only once for a shared topology and batch size. This stage builds the Jacobian sparsity pattern, reusable block-aware indexing metadata, and cuDSS LU analysis state \cite{nvidia_cudss}. Thereafter, custom CUDA kernels update the values of the base Jacobian $\bm J$, block-aware indexing avoids memory reallocation, and cuDSS refactorization accelerates linear equation solving during repeated NR iterations and training loops.

\subsubsection{BDE Jacobian Layout and In-Place Value-Update}

During the one-time symbolic analysis stage, SABLE exploits the fact that the
sparsity of the PF base Jacobian $\bm J$ is determined primarily by the nonzero
structure of $Y_{\mathrm{bus}}$. It first constructs a single-scenario sparse
base Jacobian template matching \eqref{eq:J_base_block}, where the four
voltage-dependent blocks with respect to $V_m$ and $\theta$ inherit the
bus-pair sparsity of $Y_{\mathrm{bus}}$, while the four generator-injection
derivative blocks are constant selector-type blocks. This template is then
extended across the batch through BDE to form the base BDE Jacobian, as
illustrated in Fig.~\ref{fig:fig_SABLE_PA_BDE}.

SABLE stores persistent BDE-aware metadata for the voltage-dependent
kernel-updated entries in a per-scenario form and reuses them across all BDE
batch blocks through the value-buffer offset. For
these entries, $k=1,\ldots,K_{\mathrm{upd}}$, the metadata consist of the local
value-buffer position $p_k$ and derivative-block label
$\mathrm{blk}_k\in\{dP_{\theta},dP_{V_m},dQ_{\theta},dQ_{V_m}\}$. For example,
$dP_{\theta}$ denotes the $\partial\Delta\bm P/\partial\bm\theta$ block in
\eqref{eq:J_base_block}. The metadata also include the local bus pair
$(i_k,j_k)$ and the associated coefficients $(G_k,B_k)$ of $Y_{\mathrm{bus}}$.
Here, $p_k\in\{1,\ldots,K_J\}$ maps the $k$-th updated entry to its position in
the single-scenario Jacobian value array, where $K_J$ denotes nnz of this Jacobian. These metadata are constructed once and reused throughout NR iterations and training steps.

By contrast, the state-dependent quantities are refreshed at every iteration.
For each batch $b$, the real and imaginary voltage components are formed as
\[
\bm V_{\mathrm{re}}^{(b)}
=
\bm V_m^{(b)} \odot \cos\bm\theta^{(b)}, 
\qquad
\bm V_{\mathrm{im}}^{(b)}
=
\bm V_m^{(b)} \odot \sin\bm\theta^{(b)},
\]
where $\odot$ denotes elementwise multiplication. The corresponding auxiliary
current vectors are then given by
\[
\bm I_{\mathrm{re}}^{(b)}
=
\bm G\bm V_{\mathrm{re}}^{(b)}
-
\bm B\bm V_{\mathrm{im}}^{(b)},
\qquad
\bm I_{\mathrm{im}}^{(b)}
=
\bm B\bm V_{\mathrm{re}}^{(b)}
+
\bm G\bm V_{\mathrm{im}}^{(b)} .
\]
In the implementation, these per-batch vectors are row-stacked into dense GPU
matrices, and $\bm I_{\mathrm{re}}$ and $\bm I_{\mathrm{im}}$ are computed by
CuPy matrix multiplication before each value-update kernel launch. This avoids
repeated buswise summations inside the kernel \cite{okuta2017cupy}.

Using the cached metadata and refreshed state-dependent arrays,
Algorithm~\ref{alg:kernel_fill} updates the voltage-dependent entries of the
BDE Jacobian value buffer in parallel. For each batch-entry pair $(b,k)$, one
GPU thread reads the $k$-th metadata record and uses
$(p,\mathrm{blk},i,j,G,B)$ as thread-local aliases. It also reads the corresponding local state values, including $\theta_i$ and $\theta_j$; in the formulas, $\theta_{ij}$ denotes the local angle difference
$\theta_i-\theta_j$ for the cached bus pair $(i,j)$. The thread then
evaluates the Jacobian-entry formula determined by $\mathrm{blk}$ and $(i,j)$,
and writes the result to the BDE value-buffer position $\pi$. Thus, the
kernel-updated entries across all batch blocks of the BDE Jacobian are refreshed
concurrently without reallocating memory or rebuilding the sparse structure.

\begin{figure}[t]
    \centering
    \safeincludegraphics[width=1\linewidth]{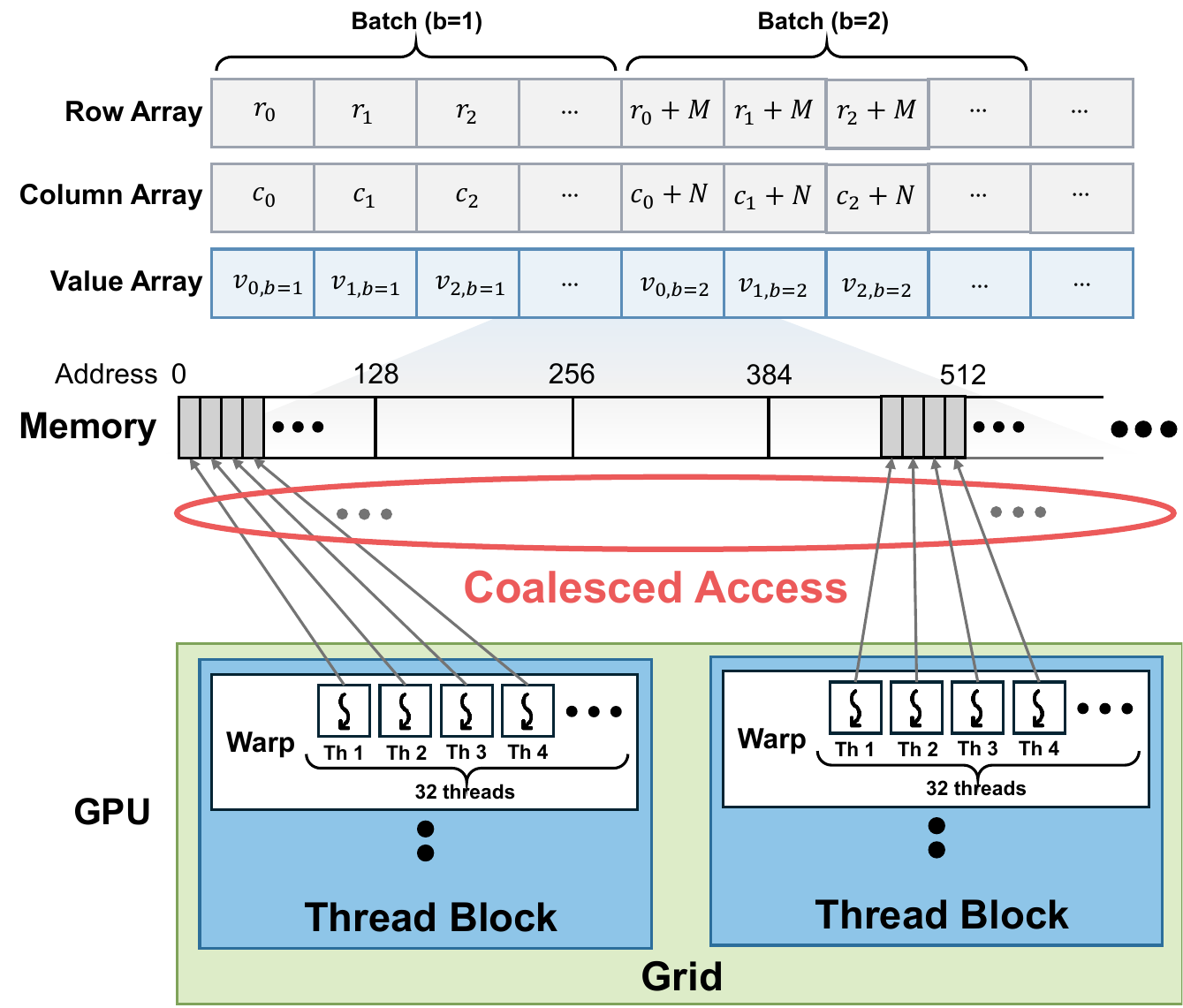}
    \caption{Illustration of coalesced memory access in the in-place value-update kernel on the BDE value array.}
    \label{fig:fig_coalescing}
\end{figure}

Because the kernel visits only the cached update entries, it skips structural
zeros and the constant generator-injection blocks, which are filled once and
then reused. When these entries are ordered by value-buffer position,
neighboring threads write to nearby locations in the linearized BDE value
buffer, promoting coalesced global-memory writes, as illustrated in
Fig.~\ref{fig:fig_coalescing}. Together, these design choices improve the throughput of BDE Jacobian value
updates while avoiding unnecessary memory reallocation.

\subsubsection{Block-Aware Indexing for Reusable Sparse Layouts}

Once the base BDE Jacobian entries are updated in place, the forward and backward passes repeatedly require the reduced BDE Jacobians, e.g., $\bm J_1$, $\bm J_1^\top$, $\bm J_2^\top$ and $\bm J_3^\top$ to be extracted from the same base layout. Simply constructing these matrices at every iteration would increase memory consumption and incur substantial overhead. SABLE avoids this cost by constructing the sparse layouts only once and refreshing only the values using a custom indexing kernel.

During the one-time symbolic analysis, SABLE first pre-expands the single-scenario indexing arrays to match the BDE structure. SABLE then selects the rows and columns required for each Jacobian operator $\bm J^{(\alpha)}$ (either $\bm J_1$, $\bm J_1^\top$, $\bm J_2^\top$ and $\bm J_3^\top$) and stores them as its cached sparse layout. Here, $\alpha$ indexes the required Jacobian operators. For each
$\bm J^{(\alpha)}$, SABLE constructs a position map
$\texttt{pos}^{(\alpha)}$ from the base BDE Jacobian value array to the corresponding reduced value array. Specifically, for the $t$-th base nonzero value $\bm{val}[t]$, $\texttt{pos}^{(\alpha)}[t]=p$ means that this entry is used as the $p$-th value of ${\bm{val}}^{(\alpha)}$, whereas a negative value
means that the base entry is not used. Accordingly, during subsequent NR and training iterations, the block-aware indexing kernel in Algorithm~\ref{alg:kernel_slice}
efficiently refreshes each $\bm J^{(\alpha)}$ by updating only its value array in parallel on the GPU according to $\texttt{pos}^{(\alpha)}$.

\begin{algorithm}[t]
\caption{Block-Aware Indexing Kernel}
\label{alg:kernel_slice}
\KwIn{
Base BDE value array $\bm{val}$ with $\mathrm{NNZ}_{\mathrm{base}}$ entries;
operator index $\alpha$;
position map $\texttt{pos}^{(\alpha)}$;
reusable value buffer $\bm{val}^{(\alpha)}$ for $\bm J^{(\alpha)}$
}
\KwOut{
Updated value buffer $\bm{val}^{(\alpha)}$ for $\bm J^{(\alpha)}$
}

\For{$t\leftarrow 0$ \KwTo $\mathrm{NNZ}_{\mathrm{base}}-1$
\tcp*[f]{in parallel over base BDE nonzeros}}{
$p\leftarrow \texttt{pos}^{(\alpha)}[t]$\;
\If{$p \ge 0$}{
$\bm{val}^{(\alpha)}[p]\leftarrow \bm{val}[t]$\;
}
}
\end{algorithm}

The same mechanism is used for both the reduced Jacobian $\bm J_1$ in the forward pass and the transpose-related Jacobian operators $\bm J_1^\top$, $\bm J_2^\top$ and $\bm J_3^\top$ required in the backward pass. Consequently, all required Jacobian operators are extracted directly from the base Jacobian in \eqref{eq:J_base_block} by refreshing only value buffers, without repeated memory reallocation, sparse pattern reconstruction, or explicit transpose formation.

Note that under a \textit{naive} sparse adaptation of the dense formulation, the backward pass would require three separate Jacobian constructions and another three explicit transpose operations. This incurs substantial overhead because repeatedly rebuilding sparse structures leads to repeated memory allocation. By contrast, since only the value arrays are refreshed while the sparse layouts remain fixed, SABLE avoids memory reallocation and
significantly reduces construction overhead.

\subsection{Stagewise Mixed Precision}
\label{subsec:direct_solve}

Finally, we introduce a stagewise MP strategy to accelerate the linear system solving that dominates the forward and backward PF layer runtime. This is particularly important when batched PF is embedded in a single-GPU
training loop.

\begin{table}[!t]
\caption{Stagewise MP policy in the SABLE forward/backward pipeline}
\label{tab:precision_policy}
\centering
\renewcommand{\arraystretch}{1.12}
\footnotesize
\begin{threeparttable}
\begin{tabular*}{\columnwidth}
{@{}p{0.60\columnwidth}
@{\extracolsep{\fill}}
>{\centering\arraybackslash}p{0.13\columnwidth}
>{\centering\arraybackslash}p{0.13\columnwidth}@{}}
\toprule
\multirow{2}{*}{\raisebox{-1.2ex}{\hspace{0.6em}SABLE stage}}& \multicolumn{2}{c}{Precision} \\
\cmidrule(lr){2-3}
& DP & SP \\
\midrule
Base BDE Jacobian value update stage & \checkmark &  \\
Forward sparse LU solve stage        &            & \checkmark \\
State/mismatch vector update stage   & \checkmark &  \\
Backward sparse LU solve stage       &            & \checkmark \\
Other backward operations stage      & \checkmark &  \\
\bottomrule
\end{tabular*}
\begin{tablenotes}
    \footnotesize
    \item DP denotes double precision, SP denotes single precision, and MP denotes mixed precision. The base BDE Jacobian value update stage denotes in-place Jacobian value computation on the shared BDE template.
\end{tablenotes}
\end{threeparttable}
\end{table}

To balance throughput and numerical robustness, we adopt the precision policy summarized in Table~\ref{tab:precision_policy}. 
In the forward pass, the in-place value-update kernel first computes the Jacobian values in DP (FP64) and then transfers the reduced BDE Jacobian $\bm J_1$ to the solver-side MP (FP32) buffer used by cuDSS, where the forward linear equation is solved in FP32. The correction is accumulated and the residual is recomputed in FP64 before the next NR iteration. In the backward pass, the same precision policy is followed as forward pass: the transpose linear equation which contains Jacobian $\bm J_1$ is likewise solved in FP32, while the remaining backpropagation computations are kept in FP64. In line with prior MP analyses~\cite{higham2022mixed, kelley2022newton}, this stagewise design reduces the dominant computational cost by aligning the linear equation solving stage with the higher-throughput FP32 path of modern GPUs, while retaining FP64 residual evaluation and convergence control for numerical robustness and final solution accuracy.

\section{Experimental Results}
\label{sec:results}
\subsection{Experimental Setup}

The CPU baseline experiments are conducted on a Windows desktop equipped with an AMD Ryzen 5 5600 6-Core Processor, while the GPU experiments use an NVIDIA GeForce RTX 4090 GPU with 24 GB of memory. The NR convergence tolerance is set to $10^{-8}$, and OOM denotes failure under the available GPU memory budget. SABLE is evaluated in two settings: standalone batched PF solving and training with an embedded PF layer.

For both settings, scenario data are generated by independently perturbing the active and reactive power demands around the nominal operating point:
\begin{equation}
\begin{aligned}
\tilde{P}_d &= P_d^{0} \odot (1+\epsilon_P),\\
\tilde{Q}_d &= Q_d^{0} \odot (1+\epsilon_Q), \qquad
\epsilon_P,\epsilon_Q \sim \mathcal{U}[-0.1,0.1],
\end{aligned}
\label{eq:pdqd_perturb}
\end{equation}
where $P_d^{0}$ and $Q_d^{0}$ denote the nominal active and reactive demand vectors of the case, and $\epsilon_P$ and $\epsilon_Q$ are sampled independently for each scenario and load. Thus, each load varies within
$\pm 10\%$ of its nominal value. For each case and evaluation setting, the same pre-generated scenario set is
used for fair comparison. The following subsections describe the setting-specific setups and then report the corresponding results.

\subsubsection{Standalone PF Setup}
Standalone PF experiments benchmark SABLE against two baselines: \emph{pandapower} as the CPU baseline and \emph{ExaPF} as the GPU baseline. Although this paper primarily evaluates end-to-end training with the PF layer, these standalone experiments show that the forward pass structure is itself sufficiently strong and competitive as an independent PF solver. All solvers use the same flat start ($V_m=1.0$ p.u., $\theta=0$). The benchmark uses four \emph{pandapower} transmission-scale cases: 1354pegase, 3120sp, 9241pegase, and ACTIVSg25k.

\begin{itemize}
    \item \textbf{pandapower:}
    We use \emph{pandapower}'s numba-accelerated NR implementation as the CPU reference. As a mature open-source PF engine that is reported to outperform MATPOWER on comparable cases, it provides a strong practical baseline \cite{thurner2018pandapower, schafer2018jacobian}.

    \item \textbf{ExaPF:}
    We use \emph{ExaPF} as the GPU reference. \emph{ExaPF} is a Julia-based GPU NR PF solver in polar form; here it is run in its documented cuDSS-based direct-solver path in FP64 \cite{exapf_juliacon,exapf_docs}.
\end{itemize}
\subsubsection{PF Layer Training Setup}
The embedded PF layer experiments use DC3 and DeepLDE as baseline learning frameworks, and their SABLE-integrated versions are denoted by DC-SABLE and D-SABLE \cite{kim2025deepl, donti2021dc3}. For DC3, the internal correction dynamics is disabled during training so that the comparison isolates the effect of replacing the PF module with SABLE. The PF layer training experiments use four systems with 793, 2312, 5658, and 7336 buses. For the training-memory study, batch sizes are swept on a power-of-two grid to identify both the practical OOM boundary and the onset of throughput saturation.

\begin{table}[!t]
\caption{Standalone PF overall runtime}
\begin{tablenotes}
\item GPU entries report runtime in milliseconds with corresponding speedup over \emph{pandapower} shown in parentheses. Boldface entries indicate the largest SABLE speedup for each system. \textit{Iter.}\ denotes the number of iterations to converge, reported in the order \emph{pandapower} / \emph{ExaPF} / SABLE.
\end{tablenotes}
\label{tab:standalone_runtime}
\centering
\scriptsize
\renewcommand{\arraystretch}{1.03}
\begin{threeparttable}
\begin{tabular*}{\columnwidth}{@{\extracolsep{\fill}}lccccc@{}}
\toprule
Case & $\nBatch$ & \shortstack[c]{\emph{pandapower}\\ {[}ms{]}} & \shortstack[c]{\emph{ExaPF}\\ {[}ms{]}} & \shortstack[c]{SABLE\\ {[}ms{]}} & Iter. \\
\midrule
\multirow{7}{*}{1354} & 4   & 112.6    & 6.1 (18.6$\times$)   & 5.4 (20.9$\times$)   & 5/5/5 \\
                       & 8   & 173.8   & 7.5 (23.1$\times$)   & 5.5 (31.6$\times$)   & 5/5/5 \\
                       & 16  & 345.9   & 10.8 (32.0$\times$)  & 5.8 (59.6$\times$)   & 5/5/5 \\
                       & 32  & 676  & 22.9 (29.5$\times$)  & 6.7 (101.0$\times$)  & 5/5/5 \\
                       & 64  & 1451.5  & 31.6 (45.9$\times$)  & 8.5 (170.8$\times$)  & 5/5/5 \\
                       & 128 & 2890.4  & 54.6 (53.0$\times$)  & 13.8 (209.4$\times$) & 5/5/5 \\
                       & 256 & 5650.5  & 95.3 (59.3$\times$)  & \textbf{22.3} (\textbf{253.4$\times$}) & 5/5/5 \\
\midrule
\multirow{7}{*}{3120} & 4   & 179   & 7.3 (24.5$\times$)   & 5.7 (31.4$\times$)   & 4/4/5 \\
                       & 8   & 351.9   & 11.9 (29.6$\times$)  & 6.0 (58.7$\times$)   & 4/5/5 \\
                       & 16  & 690.4   & 15.8 (43.7$\times$)  & 7.5 (92.5$\times$)   & 4/5/5 \\
                       & 32  & 1672.2  & 27.2 (61.5$\times$)  & 11.2 (148.8$\times$) & 4/5/5 \\
                       & 64  & 2865.4  & 48.4 (59.2$\times$)  & 18.1 (157.9$\times$) & 4/5/5 \\
                       & 128 & 5753.8  & 88.7 (64.9$\times$)  & 32.9 (174.9$\times$) & 4/5/5 \\
                       & 256 & 11834.9  & 160.7 (73.6$\times$) & \textbf{57.6} (\textbf{205.3$\times$}) & 4/5/5 \\
\midrule
\multirow{7}{*}{9241} & 4   & 692.1   & 47.6 (14.5$\times$) & 10.9                             (63.5$\times$) & 7/7/7 \\
                       & 8   & 1420.6  & 59.9 (23.7$\times$)  & 13.2 (107.6$\times$) & 7/7/7 \\
                       & 16  & 2832.9  & 99.4 (28.5$\times$)  & 21.9 (129.4$\times$) & 7/7/7 \\
                       & 32  & 5886.5  & 183.4 (32.1$\times$) & 34.9 (168.7$\times$) & 7/7/7 \\
                       & 64  & 10956 & 334.9 (32.7$\times$) & 64.8 (169.1$\times$) & 7/7/7 \\
                       & 128 & 23546 & 686.6 (34.3$\times$) & 119.9 (196.4$\times$) & 7/7/7 \\
                       & 256 & 46529 & OOM                  & \textbf{226.2} (\textbf{205.7$\times$}) & 7/--/7 \\
\midrule
\multirow{7}{*}{25k}  & 4   & 1776.5  & 43.0 (41.3$\times$)  & 14.9 (119.2$\times$) & 5/5/5 \\
                       & 8   & 3796.6  & 63.9 (59.4$\times$)  & 25.1 (151.3$\times$) & 5/5/5 \\
                       & 16  & 7286.6  & 105.7 (69.0$\times$) & 37.2 (195.9$\times$) & 5/5/5 \\
                       & 32  & 14390.6 & 181.1 (79.4$\times$) & 65.8 (218.7$\times$) & 5/5/5 \\
                       & 64  & 28938.6 & 331.7 (87.2$\times$) & \textbf{124.5} (\textbf{232.4$\times$}) & 5/5/5 \\
                       & 128 & 56975.5 & 628.6 (90.6$\times$) & 253.7 (224.6$\times$) & 5/5/5 \\
                       & 256 & 112030.6 & OOM                  & 493.9 (226.8$\times$) & 5/--/5 \\
\bottomrule
\end{tabular*}
\end{threeparttable}
\end{table}

\subsection{Standalone PF Results}

\subsubsection{NR Runtime for Solving Batched PF}

Table~\ref{tab:standalone_runtime} shows that SABLE consistently outperforms both the CPU baseline (\emph{pandapower}) and the GPU baseline (\emph{ExaPF}) across all tested systems and batch sizes. The performance gap widens with the system and batch size, indicating that the proposed BDE-based sparse execution path becomes more effective for large batched workloads. SABLE reaches up to $253.4\times$ speedup over \emph{pandapower} and up to $5.7\times$ speedup over \emph{ExaPF} on shared feasible batch sizes.

Notably, SABLE remains executable at $N_{\mathrm{batch}}=256$ for 9241pegase and ACTIVSg25k, where \emph{ExaPF} encounters OOM. This result indicates that the proposed method improves not only runtime but also the effective feasible batch
range under the same GPU memory budget. Hence, SABLE is particularly suitable for massive scenario-based PF evaluation, where both throughput and memory capacity determine practical scalability.

\subsubsection{Component-wise Acceleration Analysis}

\begin{table}[!t]
\caption{Stage-level runtime breakdown at $\nBatch=64$}
\label{tab:standalone_breakdown}
\centering
\scriptsize
\renewcommand{\arraystretch}{1.04}
\begin{threeparttable}
\begin{tabular*}{\columnwidth}{@{\extracolsep{\fill}}lcccccc@{}}
\toprule
\multirow{2}{*}{\raisebox{-0.7ex}{Case}} &
\multicolumn{3}{c}{Jacobian construction [ms]} &
\multicolumn{3}{c}{Lin.\ solv. [ms]} \\
\cmidrule(lr){2-4}\cmidrule(lr){5-7}
& \emph{ExaPF} & SABLE & Gain & \emph{ExaPF} & SABLE & Gain \\
\midrule
1354 & 3.40  & 0.32 & 10.63$\times$ & 6.80  & 1.89  & 3.60$\times$ \\
3120 & 6.10  & 0.72 & 8.47$\times$  & 10.73 & 4.80  & 2.24$\times$ \\
9241 & 67.7  & 2.04 & 33.19$\times$ & 29.1  & 14.56 & 2.00$\times$ \\
25k  & 52.21 & 5.00 & 10.4$\times$  & 70.7  & 44.73 & 1.58$\times$ \\
\bottomrule
\end{tabular*}
\begin{tablenotes}
    \footnotesize
    \item \emph{Lin.\ solv.} denotes the time required for solving the linear equations within the NR iteration. \emph{Gain} represents the speedup multiplier of SABLE over \emph{ExaPF}.
\end{tablenotes}
\end{threeparttable}
\end{table}

Table~\ref{tab:standalone_breakdown} compares \emph{ExaPF} and SABLE at the two main target stages of the proposed acceleration strategy: Jacobian construction and linear equation solving. For Jacobian construction, SABLE reduces runtime by $8.5\times$ to $33.2\times$ compared with \emph{ExaPF}. This gain is mainly attributed to the fixed BDE layout and the in-place value-update kernel, which replace repeated sparse assembly with value-only updates at preallocated locations. Linear equation solving also shows consistent advantage. Although both rely on the same cuDSS sparse LU linear solver, SABLE executes the dominant numeric solving path through a stagewise mixed-precision (MP) pipeline, which improves computational efficiency at the most expensive part of the solving phase. As a result, the linear equation solving time is reduced by $1.6\times$ to $3.6\times$ compared to \emph{ExaPF}.

In summary, Tables~\ref{tab:standalone_runtime} and \ref{tab:standalone_breakdown} show that SABLE is outstanding for accelerating NR-based PF through the in-place value-update kernel and MP pipeline.

\subsubsection{Numerical Fidelity}

\begin{table}[!t]
\caption{Jacobian agreement and NR convergence fidelity}
\label{tab:accuracy_combined}
\centering
\scriptsize
\setlength{\tabcolsep}{2.6pt}
\renewcommand{\arraystretch}{1.06}
\begin{threeparttable}
\begin{tabular*}{\columnwidth}{@{\extracolsep{\fill}}lcccc@{}}
\toprule
\multirow{2}{*}{\raisebox{-0.7ex}{Case}}
& \multicolumn{2}{c}{Jacobian error} 
& \multicolumn{2}{c}{NR} \\
\cmidrule(lr){2-3}\cmidrule(lr){4-5}
& Max abs. error & Rel. error at max 
& Iter. & Final mismatch \\
\midrule
1354 & $3.53{\times}10^{-10}$ & $2.07{\times}10^{-14}$ & 5/5 & $5.77{\times}10^{-12}$ / $1.42{\times}10^{-11}$ \\
3120 & $3.93{\times}10^{-9}$  & $1.74{\times}10^{-13}$ & 4/5 & $6.80{\times}10^{-10}$ / $1.49{\times}10^{-11}$ \\
9241 & $2.00{\times}10^{-9}$  & $2.19{\times}10^{-13}$ & 7/7 & $1.23{\times}10^{-11}$ / $8.88{\times}10^{-11}$ \\
25k  & $5.60{\times}10^{-9}$  & $8.51{\times}10^{-14}$ & 5/5 & $2.16{\times}10^{-11}$ / $2.38{\times}10^{-10}$ \\
\bottomrule
\end{tabular*}
\end{threeparttable}
\end{table}

Table~\ref{tab:accuracy_combined} evaluates numerical fidelity at
$\nBatch=64$ by comparing the Jacobian mismatch between SABLE and the CPU
\emph{pandapower} reference and by reporting NR convergence in DP/MP order. Max abs. error and Rel. error at max denote the maximum absolute
entry-wise error and the corresponding relative error, respectively. The NR
columns report Iter. and Final mismatch, where Iter. denotes the number of
iterations required for convergence and Final mismatch denotes the final $\|\bm F\|_{\infty}$.

The proposed Jacobian update shows high agreement with the reference, with
maximum absolute and relative errors below $5.60\times10^{-9}$ and
$2.19\times10^{-13}$, respectively. MP also preserves NR convergence, requiring
the same or at most one additional iteration compared with DP and satisfying
the prescribed mismatch tolerance in all tested cases.

\subsection{PF Layer Results}
\begin{figure}[!t]
    \centering
    \safeincludegraphics[width=\columnwidth]{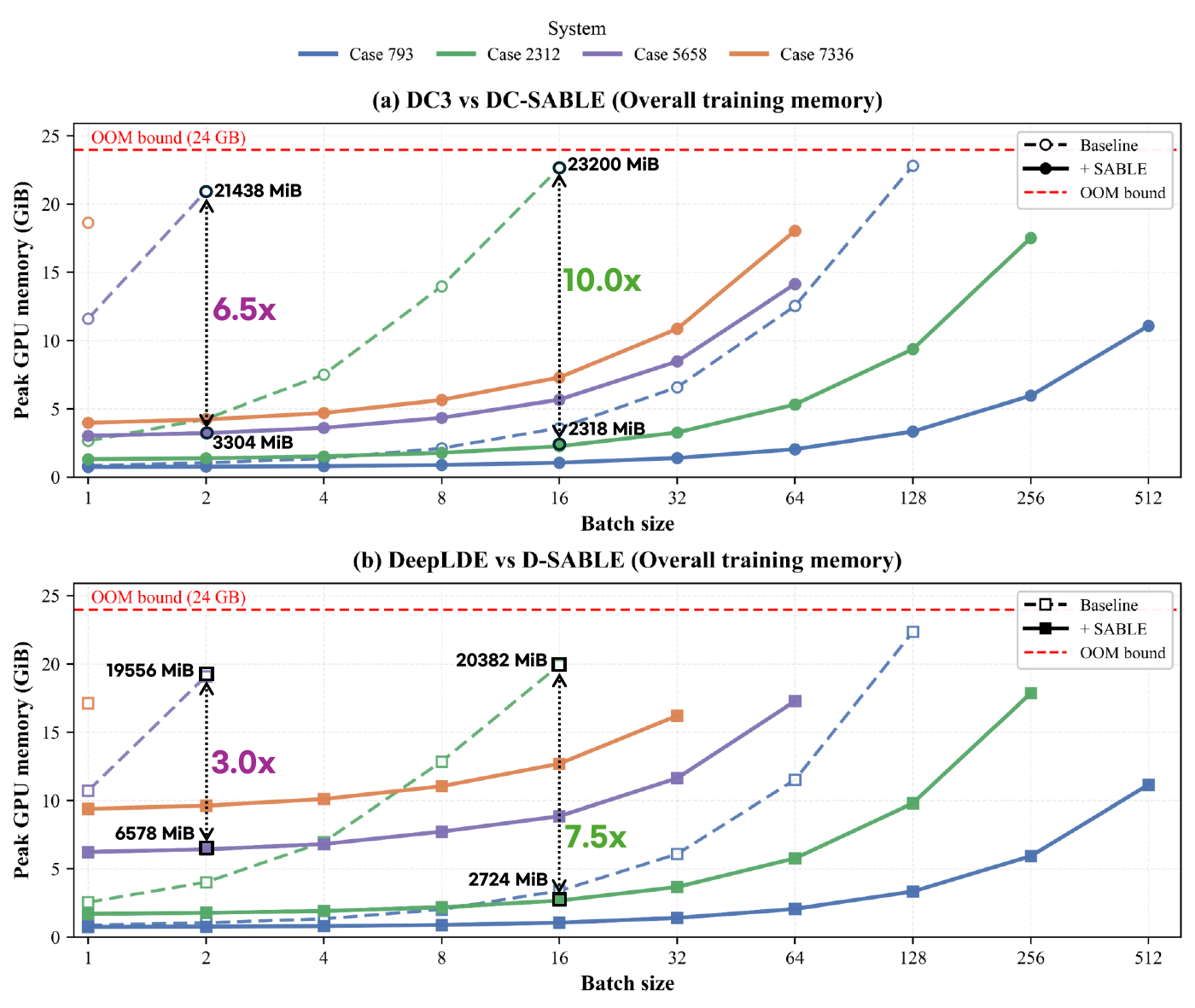}
    \caption{Peak training memory versus batch size for baselines and their SABLE-integrated models.}
    \label{fig:solver_memory}
\end{figure}

\begin{table}[!t]
    \caption{Maximum Allowable Batch Size Improvements by SABLE}
    \label{tab:max_batch_size}
    \centering
    \scriptsize
    \renewcommand{\arraystretch}{1.04}
    \begin{tabular*}{\columnwidth}{@{\extracolsep{\fill}}lcc@{}}
    \toprule
    Case &
    \shortstack[c]{DC3 $\rightarrow$ DC-SABLE} &
    \shortstack[c]{DeepLDE $\rightarrow$ D-SABLE} \\
    \midrule
    793  & 128 $\rightarrow$ 512 (\textbf{4$\times$})  & 128 $\rightarrow$ 512 (\textbf{4$\times$}) \\
    2312 & 16 $\rightarrow$ 256 (\textbf{16$\times$})  & 16 $\rightarrow$ 256 (\textbf{16$\times$}) \\
    5658 & 2 $\rightarrow$ 64 (\textbf{32$\times$})   & 2 $\rightarrow$ 64 (\textbf{32$\times$}) \\
    7336 & 1 $\rightarrow$ 64 (\textbf{64$\times$})   & 1 $\rightarrow$ 32 (\textbf{32$\times$}) \\
    \bottomrule
    \end{tabular*}
    \begin{tablenotes}
    \item Maximum trainable batch sizes are reported as baseline $\rightarrow$ SABLE-integrated model; the increase factors are shown in parentheses.
    \end{tablenotes}
\end{table}

\subsubsection{Training-Memory Scaling}
Fig.~\ref{fig:solver_memory} compares GPU memory usage across batch sizes for DC3 and DeepLDE with and without SABLE. Baseline models show rapidly increasing memory usage as batch size grows, whereas SABLE-integrated models grow more slowly. This difference becomes more pronounced for larger systems,
where sparsity preservation has greater effect on the total training memory footprint. At the largest trainable batch sizes, SABLE improves memory efficiency by about
$6.5\times$ for DC3 and $3.0\times$ for DeepLDE on the 2312-bus system. On the
5658-bus system, the corresponding gains increase to about $10.0\times$ and
$7.5\times$, respectively.

Table~\ref{tab:max_batch_size} summarizes maximum feasible batch size for each system. SABLE increases the maximum allowable batch size by up to $64\times$ for DC3 and $32\times$ for DeepLDE. This memory efficiency improvement comes not from the surrounding learning architecture, but directly from the proposed PF layer implementation. The proposed design preserves power-system sparsity through end-to-end training and avoids unnecessary memory reallocation, leading to a highly efficient computation structure. As observed in Fig.~\ref{fig:solver_memory}, SABLE fully exploits system sparsity, so memory usage grows slowly as batch size increases. Therefore, on GPUs with larger VRAM, SABLE's memory efficiency gain is expected to become even greater.

\begin{figure}[!t]
    \centering
    \safeincludegraphics[width=\columnwidth]{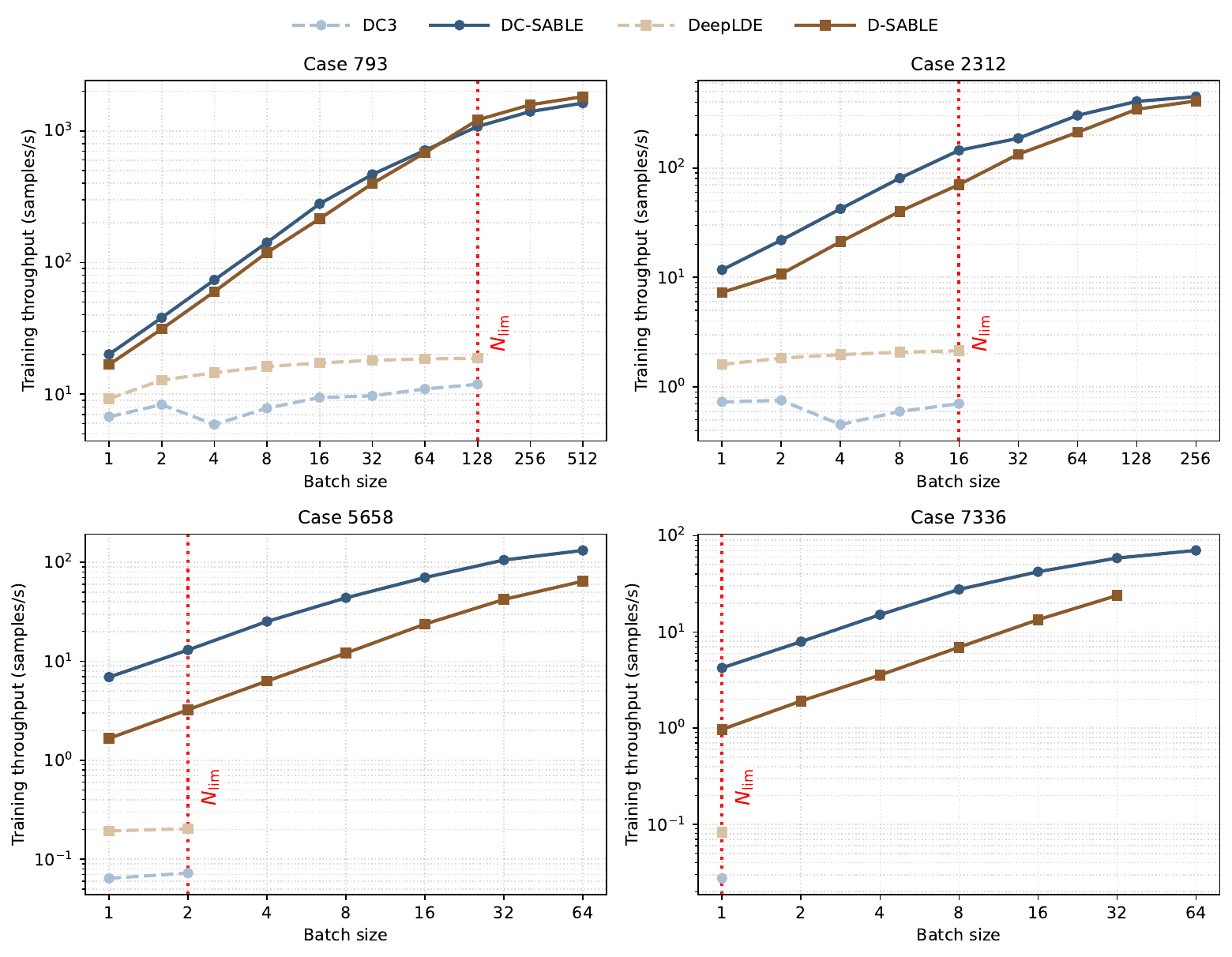}
    \caption{End-to-end training throughput with a PF layer versus batch size. The red vertical line marks $\nLim$.}
    \label{fig:sitl_throughput_scaling}
\end{figure}
\begin{table}[!t]
\caption{PF layer throughput at $\nLim$}
\label{tab:sitl_throughput}
\centering
\scriptsize
\setlength{\tabcolsep}{3.0pt}
\renewcommand{\arraystretch}{1.08}
\begin{threeparttable}
\resizebox{\columnwidth}{!}{%
\begin{tabular}{c c c c c c c}
\toprule
\multirow{2}{*}{\raisebox{-0.7ex}{Case}} & \multirow{2}{*}{\raisebox{-0.7ex}{$\nLim$}} & \multirow{2}{*}{\raisebox{-0.7ex}{Base}}
& \multicolumn{2}{c}{Training Throughput}
& \multicolumn{2}{c}{PF Fwd+Bwd Throughput} \\
\cmidrule(lr){4-5}\cmidrule(lr){6-7}
& & & Baseline & with SABLE & Baseline & with SABLE \\
\midrule
793  & 128 & DC3      & 11.91 & 1079.26 (\textbf{90.6$\times$}) & 12.05 & 1147.26 (\textbf{95.2$\times$}) \\
     &     & DeepLDE  & 18.76 & 1210.52 (\textbf{64.5$\times$}) & 18.83 & 1365.77 (\textbf{72.5$\times$}) \\
\midrule
2312 & 16  & DC3      & 0.70  & 144.67 (\textbf{206.7$\times$}) & 0.71  & 152.98 (\textbf{215.5$\times$}) \\
     &     & DeepLDE  & 2.14  & 70.43 (\textbf{32.9$\times$})   & 2.16  & 139.75 (\textbf{64.6$\times$}) \\
\midrule
5658 & 2   & DC3      & 0.072 & 13.04 (\textbf{181.1$\times$})  & 0.073 & 14.05 (\textbf{192.5$\times$}) \\
     &     & DeepLDE  & 0.20  & 3.24 (\textbf{16.2$\times$})    & 0.21  & 13.94 (\textbf{66.4$\times$}) \\
\midrule
7336 & 1   & DC3      & 0.028 & 4.23 (\textbf{151.1$\times$})   & 0.028 & 4.47 (\textbf{159.6$\times$}) \\
     &     & DeepLDE  & 0.083 & 0.97 (\textbf{11.7$\times$})    & 0.088 & 3.78 (\textbf{43.0$\times$}) \\
\bottomrule
\end{tabular}%
}
\begin{tablenotes}
    \item \textit{Training Throughput} denotes end-to-end training throughput in samples/s, 
    \\ \textit{PF Fwd+Bwd Throughput} includes only PF layer forward/backward passes. Throughput ratios relative to the corresponding baseline are shown in parentheses.
\end{tablenotes}
\end{threeparttable}
\end{table}

\subsubsection{Throughput at the Baseline-Feasible Batch}
Fig.~\ref{fig:sitl_throughput_scaling} and Table~\ref{tab:sitl_throughput} reveal two key throughput observations, measured in samples/s at \(\nLim\), the largest sampled batch size that the baseline can execute without OOM. First, as batch size increases, baseline training throughput rapidly saturates, whereas SABLE continues to scale and saturates only at much larger batch sizes. This indicates that SABLE maintains higher GPU utilization and can process larger batches efficiently. This trend is specially pronounced for the 5658- and 7336-bus systems. For these cases, GPUs with larger VRAM would further increase the maximum comparable batch size, suggesting an even larger throughput gap between SABLE and the baseline. Table~\ref{tab:sitl_throughput} shows that SABLE improves PF forward/backward and end-to-end training throughput by up to $215.5\times$ and $206.7\times$ for DC3, and by up to $72.5\times$ and $64.5\times$ for DeepLDE, respectively. These throughput results show that SABLE substantially reduces the training
cost of physics-embedded learning models. 

\subsubsection{Representative Training Convergence}

\begin{figure}[!t]
\centering
\safeincludegraphics[width=1.0\columnwidth]{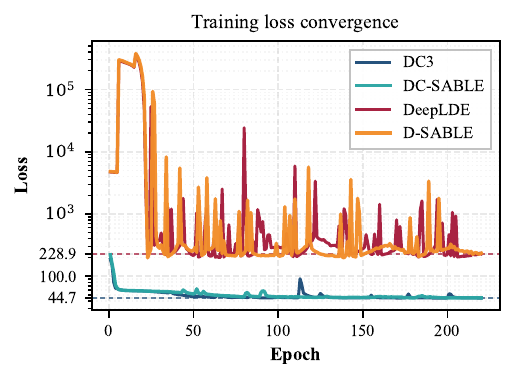} 
\caption{Representative training-loss trajectories versus epoch for baseline and SABLE embedded models.}
\label{fig:sitl_training_curve}
\end{figure}

Fig.~\ref{fig:sitl_training_curve} shows that, within each learning framework,
the baseline and SABLE-embedded models achieve nearly identical
loss-convergence behavior. The loss-scale difference between DC3 and DeepLDE reflects their different constraint formulations, including thermal
limit treatment, rather than the use of SABLE. This confirms that SABLE does not introduce
observable degradation in training dynamics, thereby supporting numerical stability of the proposed PF layer implementation during end-to-end
learning.

Overall, these results demonstrate that SABLE addresses two major
limitations of physics-embedded learning models for power-system
optimization: excessive memory consumption and long training time. The
proposed framework therefore provides a practically scalable and
computationally efficient training structure for large-scale power-system
applications.

\section{Conclusion}
This paper presented SABLE, a GPU-based sparse batched PF accelerator for standalone PF evaluation and differentiable PF layers. By reusing a shared sparse-pattern execution structure, SABLE supports both solving batched PF in the forward pass and differentiable backward propagation within GPU-resident training pipelines. The results indicate that SABLE remains competitive as a standalone solver while substantially expanding the feasible training regime and improving end-to-end training throughput for large systems with embedded PF. More broadly, by reducing embedded PF overhead, SABLE can be applied to power-system problems requiring repeated batched PF solving and differentiation within larger optimization or learning loops.

\end{document}